\documentstyle[prc,preprint,aps]{revtex}
\begin{document}
\tighten
\draft
\title{Signal for supernova $\nu_\mu$ and $\nu_\tau$ neutrinos in
water \v{C}erenkov detectors}

\author{K. Langanke$^1$, P. Vogel$^2$ and E. Kolbe$^3$}
\address{
$^1$ W. K. Kellogg Radiation Laboratory, 106-38, California
Institute of Technology\\ Pasadena, California 91125 USA \\
$^2$ Physics Department, California
Institute of Technology\\ Pasadena, California 91125 USA \\
$^3$ Institut f\"ur Physik, Universit\"at Basel, Switzerland}
\date{\today}
\maketitle

\begin{abstract}
We suggest that photons with energies between 5 and 10 MeV,
generated by the ($\nu,\nu'p\gamma$) and
($\nu,\nu'n\gamma$) reactions on $^{16}$O, constitute a signal
which allows a unique identification of supernova
$\nu_\mu$ and $\nu_\tau$ neutrinos in water \v{C}erenkov detectors. We
calculate the yield of such $\gamma$ events and estimate that a
few hundred of them
would be detected in Superkamiokande for a supernova at 10 kpc distance.
\end{abstract}

\pagestyle{plain}
\newpage
Neutrinos play a decisive role in various
stages of supernova evolution \cite{Bethe}.
In particular, the gravitational binding energy
of the nascent neutron star is released by neutrino-pair production
\cite{Cooperstein}.
It is the neutrinos generated
during this cooling and deleptonization phase of the hot remnant
core which will be mainly observed in earthbound detectors.
Although pairs of all three flavors are generated with equal
luminosity \cite{Qian},
due to their smaller opacities
$\nu_{\mu}$ and $\nu_{\tau}$ neutrinos and their antiparticles decouple
at smaller radii, and thus higher temperatures in the core, than
$\nu_e$ and ${\bar \nu_e}$ neutrinos.
As the neutrinos decouple in neutron-rich
matter, which is less transparent for $\nu_e$ than for ${\bar \nu_e}$, it is
expected on general grounds that the neutrino spectra after decoupling
obey the temperature hierarchy \cite{Qian},
$T_{\nu_x} > T_{\bar \nu_e} > T_{\nu_e}$,
where $\nu_x$ stands for $\nu_\mu$, $\nu_\tau$ and their antiparticles,
which are assumed to have identical spectra. The neutrino spectra can be
approximately described by Fermi-Dirac (FD) distributions with zero chemical
potential and
$T_{\nu_x}=8$ MeV, $ T_{\bar \nu_e}=5$ MeV, and $ T_{\nu_e}=3.5$ MeV,
corresponding to average neutrino energies of
$\langle E_{\nu_x} \rangle =25$ MeV,
$\langle E_{\bar \nu_e} \rangle =16$ MeV, and
$\langle E_{\nu_e} \rangle =11$ MeV. More elaborate investigations
of neutrino production in supernovae indicate that the high-energy
tail of the neutrino spectra is better described by a Fermi-Dirac
distribution with a finite chemical potential \cite{Janka,Yong}.

In what is considered the birth of neutrino astrophysics,
neutrinos from supernova SN1987A have been detected by the Kamiokande
\cite{Kamio} and IMB \cite{IMB} water \v{C}erenkov detectors (11 and 8 events,
respectively).
It is generally assumed that these events originated
from the
${\bar \nu}_e +p \rightarrow n+e^+$ reaction in water. The detection of
$\nu_e$ and
$\nu_x$ neutrinos via the
$\nu+e\rightarrow \nu'+e'$ scattering
or the
$^{16}$O($\nu_e,e^-$)$^{16}$F reaction
was strongly suppressed by the small effective cross sections
of these processes. The observability of supernova
neutrinos will significantly improve
when the Superkamiokande (SK) detector becomes operational \cite{Totsuka}.
This detector, with about 15 times the target material for supernova
neutrinos of Kamiokande
and a lower threshold of $E_{\rm th} = 5$ MeV, will be capable to
detect also the recoil electrons from
$\nu+e\rightarrow \nu'+e'$. In principle,  $\nu_x$ induced
neutrino-electron scattering events can be
separated from everything else in SK using their angular
distributions and energy spectra \cite{Totsuka}.
However, only about one third of the $\nu_x + e$ scattering events
will have energies distinctly larger than the recoil electrons from
$\nu_e + e$ and ${\bar \nu}_e + e$ scattering. Moreover, these
higher energy electron recoils have to be separated by their
direction from the much more numerous positrons from
${\bar \nu}_e + p \rightarrow n + e^+$ with the same energy.

In this Letter we suggest another signal in water \v{C}erenkov detectors
which allows one to unambigously identify $\nu_x$ induced events.
The basis of our proposal is the fact that SK can observe photons
with energies larger than 5 MeV \cite{Koshiba}.
Schematically our detection scheme works
as follows (Fig. 1). Supernova $\nu_x$ neutrinos, with average energies
of $\approx 25$ MeV, will predominantly excite $1^-$ and $2^-$ giant
resonances in $^{16}$O via  the
$^{16}$O($\nu_x,\nu^\prime_x$)$^{16}$O$^\star$ neutral current reaction
\cite{Kolbe92}. These resonances are above the particle thresholds and
will mainly decay by proton and neutron emission. (Decay into the
$\alpha$ channel, although energetically allowed, is strongly suppressed
by isospin conservation \cite{Kolbe92}). Although the proton and
neutron decays will be mainly to the ground states of
$^{15}$N and $^{15}$O, respectively, some of these decays will go to
excited states in these nuclei. If these excited states are below
the particle thresholds in
$^{15}$N ($E^\star < 10.2$ MeV) or
$^{15}$O ($E^\star < 7.3$ MeV),
they will decay by $\gamma$ emission.
As the first excited states in both nuclei
($E^\star =5.27$ MeV in $^{15}$N and
$E^\star = 5.18$ MeV in $^{15}$O)
are at energies larger than the SK detection threshold, all of the excited
states in $^{15}$N and $^{15}$O below the respective particle
thresholds will emit photons which can be observed in SK.

Of course, it is important to estimate the effective
$^{16}$O($\nu_x,\nu^\prime_x p \gamma$) and
$^{16}$O($\nu_x,\nu^\prime_x n \gamma$)  cross sections for SK and to
compare it to the effective ``background'' cross sections, stemming from the
${\bar \nu}_e +p \rightarrow n+e^+$ and $\nu+e\rightarrow \nu'+e'$ events
with energy release similar to the energy of the photons.
Assuming equal luminosities for all neutrino species ($i=\nu_e, {\bar \nu}_e,
\nu_x$) leaving a supernova, the relative event rate $(\sigma_{\rm eff})$
for a specific neutrino-induced process
in a water \v{C}erenkov detector is \cite{Qian}
\begin{equation}
(\sigma_{\rm eff}) \sim
\frac{n}{\langle E \rangle}
\int dE f(E) \sigma (E)
\end{equation}
where $f$ is the neutrino energy spectrum and
the factor $1/\langle E \rangle$
accounts for the ratio of fluxes for the different neutrino flavors.
$\sigma (E)$ is the total cross section
for the neutrino-induced process
and $n$ is the number of targets for an individual neutrino process
in a single water molecule ($n=10$ for neutrino-electron scattering,
$n=2$ for
${\bar \nu}_e +p \rightarrow n+e^+$ and $n=1$ for neutrino reactions
on $^{16}$O).

To calculate the
$^{16}$O($\nu_x,\nu^\prime_x p \gamma$) and
$^{16}$O($\nu_x,\nu^\prime_x n \gamma$) cross sections we assume
a two-step process.
In the first step we calculate
the
$^{16}$O($\nu_x,\nu^\prime_x$)$^{16}$O$^{*}$
cross section as a function of excitation energy in $^{16}$O within
the continuum random phase approximation (CRPA). In the second step
we calculate for each final state
with well-defined energy, angular momentum and parity
the branching ratios into the various decay channels
using the statistical model code SMOKER \cite{physrep},
considering proton, neutron, $\alpha$ and $\gamma$ emission. As possible
final states in the residual nucleus the SMOKER code considers
the experimentally known levels supplemented at higher energies
by an appropriate level density formula \cite{physrep}. If the decay leads
to an excited level of the residual nucleus (e.g. to p+$^{15}$N$^*$), we
calculate the branching ratios for the decay of this level in an
analogous way. Keeping track of the energies of the ejected particles and
photons during the cascade, and weighting them with appropriate
branching ratios and the corresponding differential
$^{16}$O($\nu_x,\nu^\prime_x$)$^{16}$O$^{*}$ cross section, we determine
the various particle (proton, neutron, $\alpha$) and photon spectra for
the ($\nu_x,\nu^\prime_x$) reaction on $^{16}$O. We performed a similar
calculation also for the
(${\bar \nu_x}, {\bar \nu}^\prime_x$) reaction on $^{16}$O.
The contribution of various neutrino energies was
weighted according to a (normalized) distribution $f(E)$.
Note that
the same CRPA approach
has been successfully
applied to the muon capture on $^{16}$O \cite{Kolbe94b}.
The model is described in details in Refs. \cite{Krewald,Kolbe92}.
As residual interaction we adopt the finite-range force based on the
Bonn potential \cite{Nakayama}.
A similar two-step approach (combining CRPA and statistical model)
has been tested successfully against the integrated
$(\gamma p)/ (\gamma n)$
data on $^{16}$O \cite{Kolbe92}.

The total and partial cross sections
for $\nu_x$ and
${\bar \nu_x}$ induced neutral current reactions on $^{16}$O
were evaluated first using the
Fermi-Dirac neutrino spectrum with zero chemical potential $\mu$
and temperature $T=8$ MeV (FD1) \cite{Woosley}. The results are listed
in Table I. The total
($\nu_x, \nu^\prime_x$) and
(${\bar \nu_x}, {\bar \nu}^\prime_x$) cross sections
are roughly the same as the
vector-axial vector interference term is rather unimportant. As expected,
the partial cross sections for decay into proton and neutron channels
dominate the total cross section. The proton channel is favored
over the neutron channel by the lower threshold in $^{16}$O.
We find that a significant fraction of these decays go to
excited states in $^{15}$N and $^{15}$O below particle thresholds
and thus decay by $\gamma$ emission. The relatively larger importance
of this decay mode in $^{15}$N ($\approx 24\%$) than in $^{15}$O
($\approx 6\%$) reflects the larger number of final states
in $^{15}$N due to the higher particle threshold. We find $3.2 \cdot
10^{-42}$ cm$^2$ for the total $\gamma$ producing
cross section for each flavor of  $\nu_x$ plus ${\bar \nu}_x$
in the neutral current  reactions on $^{16}$O,
obtained by adding the ($p\gamma$) and $(n\gamma$) partial cross sections.

As discussed above, the $\nu_x$ and ${\bar \nu}_x$ induced reactions
will produce $\gamma$ events in the energy range $E\approx 5-10$ MeV.
The other events at these energies (a ``background" for our purpose)
 will stem mainly from the
${\bar \nu_e} + p \rightarrow n + e^+$ reaction.
Adopting the Fermi-Dirac distribution with $T=5$ MeV
and zero chemical potential, we calculate a total cross section for this
reaction of $47 \cdot 10^{-42}$ cm$^2$. (This includes the factor $n=2$
for the two protons in a water molecule. The result is somewhat smaller
than that quoted in \cite{Qian} where minor effects, such as the
weak magnetism and recoil were not included.) However, the energy spectrum
of positrons as seen by SK is peaked at around 15 MeV and only a small
fraction of events are in the energy window $E=5-10$ MeV. This becomes obvious
in Fig. 2, where we compare the positron spectrum with the $\gamma$ spectrum
calculated for the $\nu_x$ and ${\bar \nu}_x$ induced reaction on $^{16}$O.
The latter has been multiplied by a factor 2 (to account for $\nu_\mu$
and $\nu_\tau$ neutrinos) and by 16/25 to consider the ratio
of ${\bar \nu_e}$ and $\nu_x$ fluxes ($\sim \langle E_{\nu_x} \rangle /
\langle E_{{\bar \nu}_e} \rangle$, see Eq. (1))) at the detector.
An energy resolution 14$\%$/($E/10)^{1/2}$ \cite{Koshiba},
where $E$ is in MeV, i.e., 1 MeV
for the energies of interest, has been assumed for the detector.
As is obvious from Fig. 2, the $\gamma$ spectrum constitutes a clear
signal at $E=5-7$ MeV on top of a smooth background from
the ${\bar \nu_e} + p \rightarrow n + e^+$ reaction.
Our calculation predicts most of the photons to stem from the decay
of the 3 lowest levels in $^{15}$N and $^{15}$O.
Further, in Fig. 2 we assume that the detector will record all
photons in a possible
cascade of several $\gamma$ rays as a single event. Let us stress
that each of such multi-photon events will contain at least one photon
above the 5 MeV threshold.

Other possible backgrounds are
neutrino-electron scattering and charged current reactions on $^{16}$O.
For these reactions we find smooth electron or positron spectra,
whose cross sections in the
interval $E=5-10$ MeV (normalized with the appropriate flux ratios
and target numbers $n$) are much smaller than the $\gamma$ signal.
Water also contains a tiny amount of $^{18}$O and even less $^{17}$O.
However, their
natural abundances ($\approx 0.2\%$ and $0.04\%$, respectively)
are too small for neutrino reactions
on $^{18}$O to be of importance (see Ref. \cite{Haxton} for the calculated
charged current cross sections).

We then repeated our calculation of the $\nu_x$ and ${\bar \nu}_x$
induced reactions on $^{16}$O, using a Fermi-Dirac neutrino spectrum
(FD2) with $T=6.26$ MeV and $\mu=3T$ \cite{Yong},
\begin{equation}
f(E) ~\sim~ \frac{E^2}{1 + {\rm exp}[(E-\mu)/T]} ~,
\end{equation}
which has the same
average neutrino energy as the FD1 distribution.
We find that the FD2 total and partial cross sections are smaller by about
a factor of two when compared to the FD1 results (see Table 1).
Noting that the main contribution to the cross sections comes from
neutrinos with $E_\nu > \langle E_\nu \rangle$, this scaling simply reflects
the ratio of the two Fermi-Dirac distributions in that energy region.
At the same time, the cross section for the dominant reaction
${\bar \nu_e} + p \rightarrow n + e^+$ is changed only slightly,
to $44 \cdot 10^{-42}$ cm$^2$ when the FD spectrum with $T=5$ MeV and $\mu=0$
 is replaced by a
spectrum with $T=4$ MeV and $\mu=3T$.

We note that photodissociation data confirm a significant decay rate
of the giant dipole resonance in $^{16}$O by proton and neutron emission
into excited states of $^{15}$N and $^{15}$O \cite{Ajzenberg}.
In agreement with our model these decays mainly lead to the first 3 excited
levels in these nuclei and are relatively larger in $^{15}$N than in
$^{15}$O
\cite{Caldwell,Horowitz}. While the total decay rate
appears to be in reasonable agreement with our calculation, the data
suggest a preference of the decay to the $3/2^-$ state at $\approx 6.3$ MeV
over the decay to the positive-parity states at around 5.3 MeV,
caused by nuclear structure effects beyond our present model
\cite{Caldwell,Horowitz}. This suggests that a fraction of the signal,
predicted by our calculation at $E \approx 5.3$ MeV just above the
SK detection threshold, is to be shifted to 6.3 MeV, where it can be easier
detected in SK. To clarify this point, a  detailed investigation
of the photodissociation process on $^{16}$O within the
current model is in progress.

Superkamiokande is expected to detect about 4000 positrons from the
${\bar \nu_e} + p \rightarrow n + e^+$ reaction \cite{Totsuka}
for a supernova going off at 10 kpc ($\approx 3 \cdot 10^4$ lightyears
or the distance to the galactic center).
By scaling the respective effective cross sections, we estimate
that such a supernova will produce about 360 (FD1) or 190 (FD2)
$\gamma$ events in the energy window $E=5-10$ MeV, to be compared
with a smooth background of about 270 positron events from the
 ${\bar \nu_e} + p \rightarrow n + e^+$ reaction in the same energy window.
This number of
events produced by supernova $\nu_x$ neutrinos via the scheme proposed here
is larger than the total number of events
expected from $\nu_x$-electron scattering (about 80
events \cite{Totsuka}). More importantly, the $\gamma$ signal
can be unambiguously identified from the observed spectrum in the SK detector,
in contrast to the more difficult identification from $\nu_x$-electron
scattering.

In conclusion, we propose a novel signal for the identification
of supernova $\mu$ and $\tau$ neutrinos in water \v{C}erenkov detectors.
Our suggestion is based on the fact that the levels in $^{16}$O,
which are excited by inelastic neutral current scattering of
supernova $\nu_x$ neutrinos,
have decay branches via proton and neutron emission into
excited states in $^{15}$N and $^{15}$O. These states, in turn,
decay by emission of photons with $E_\gamma > 5$ MeV, which can be detected
in the Superkamiokande water \v{C}erenkov detector. We show that
the expected number of $\gamma$ events in the relevant energy window
$E =5-10$ MeV is noticeably larger than the positron or electron background
expected from other neutrino reactions in water.
It is amusing to note that the $\nu_x$ neutrinos from
SN1987A at 50 kpc would have
created about 10 photons in SK (which did not exist at that
time, unfortunately), the same number of events as all of the recorded
events in the Kamiokande or IMB detectors then, which launched the
era of neutrino astronomy.

We are grateful to F.-K. Thielemann for providing us with his statistical
model code SMOKER.
We thank Y.-Z. Qian and F.-K. Thielemann
for helpful discussions.
This work was supported by  the U.S. National Science Foundation
(grants  PHY94-12818 and PHY94-20470), by the U.S. Department
of Energy (DE-FG03-88ER-40397) and by the Swiss Nationalfonds.

\newpage

\begin{table}[btph]
\caption[Table 1]{Total and partial cross sections for $\nu_x$ and
${\bar \nu}_x$ induced reactions on $^{16}$O, calculated for
a Fermi-Dirac neutrino spectrum with temperature and
chemical potential
($T$=8 MeV, $\mu=0$) (upper part) and
($T$=6.26 MeV, $\mu=3T$) (lower part).
}
\begin{center}
\begin{tabular}{|l|r|l|r|} \hline \hline
 reaction & $\sigma_{tot}$  [$10^{-42}cm^2$] &
 reaction & $\sigma_{tot}$  [$10^{-42}cm^2$]
\\ \hline\hline
$^{16}$O($\nu_x,\nu^\prime_x$)X & 5.90 &
$^{16}$O(${\bar \nu_x}, {\bar \nu^\prime_x}$)X & 4.48 \\
$^{16}$O($\nu_x,\nu^\prime_x p$)$^{15}$N &   3.75   &
$^{16}$O(${\bar \nu_x}, {\bar \nu^\prime_x} p$)$^{15}$N &  2.93 \\
$^{16}$O($\nu_x,\nu^\prime_x n$)$^{15}$O &   1.76   &
$^{16}$O(${\bar \nu_x}, {\bar \nu^\prime_x} n$)$^{15}$O &  1.29 \\
$^{16}$O($\nu_x,\nu^\prime_x p \gamma$)$^{15}$N &   1.41   &
$^{16}$O(${\bar \nu_x}, {\bar \nu^\prime_x} p \gamma$)$^{15}$N &  1.09 \\
$^{16}$O($\nu_x,\nu^\prime_x n \gamma$)$^{15}$O &   0.37   &
$^{16}$O(${\bar \nu_x}, {\bar \nu^\prime_x} n \gamma$)$^{15}$O &  0.28 \\
\hline
$^{16}$O($\nu_x,\nu^\prime_x$)X & 3.08 &
$^{16}$O(${\bar \nu_x}, {\bar \nu^\prime_x}$)X & 2.50 \\
$^{16}$O($\nu_x,\nu^\prime_x p$)$^{15}$N &   2.02   &
$^{16}$O(${\bar \nu_x}, {\bar \nu^\prime_x} p$)$^{15}$N &  1.69 \\
$^{16}$O($\nu_x,\nu^\prime_x n$)$^{15}$O &   0.90   &
$^{16}$O(${\bar \nu_x}, {\bar \nu^\prime_x} n$)$^{15}$O &  0.70 \\
$^{16}$O($\nu_x,\nu^\prime_x p \gamma$)$^{15}$N &   0.72   &
$^{16}$O(${\bar \nu_x}, {\bar \nu^\prime_x} p \gamma$)$^{15}$N &  0.59 \\
$^{16}$O($\nu_x,\nu^\prime_x n \gamma$)$^{15}$O &   0.18   &
$^{16}$O(${\bar \nu_x}, {\bar \nu^\prime_x} n \gamma$)$^{15}$O &  0.14 \\
\hline\hline
\end{tabular}
\end{center}
\end{table}

\newpage

\begin{figure}
\caption{
Schematic illustration of the detection scheme for supernova $\nu_\mu$
and $\nu_\tau$ neutrinos in water \v{C}erenkov detectors.}
\end{figure}

\begin{figure}
\caption{Signal expected from supernova neutrinos in a water
\v{C}erenkov detector. The solid line is the sum of
the $\gamma$ spectrum, generated by
$\nu_x$
and ${\bar \nu}_x$ reactions on $^{16}$O,
and of the positron spectrum (dashed line)
from the ${\bar \nu_e} + p \rightarrow n + e^+$ reaction.
The upper part (a) has been calculated assuming Fermi-Dirac
neutrino distributions
with ($T=8$ MeV, $\mu=0$) and ($T=5$ MeV ,$\mu=0$) for
$\nu_x$ and ${\bar \nu_e}$ neutrinos, respectively.  In the lower part (b)
Fermi-Dirac neutrino distributions with ($T=6.26$ MeV, $\mu=3T$) and
($T=4$ MeV, $\mu=3T$) have been assumed for $\nu_x$ and ${\bar \nu_e}$
neutrinos. The energy $E$ refers to the photon or
positron energy, respectively.
The spectra
are in arbitrary units.}
\end{figure}

\end{document}